\documentclass[10pt,conference]{IEEEtran}

\usepackage[utf8]{inputenc}
\usepackage[T1]{fontenc}
\usepackage[english]{babel}
\usepackage{csquotes}
\usepackage[style=ieee,sortcites=true,sorting=none,maxbibnames=6, backend=biber,mincitenames=1,maxcitenames=2, dashed=false]{biblatex}
\usepackage{graphicx}
\usepackage{amsmath}
\usepackage{amssymb}
\usepackage{amsthm}
\usepackage{mathtools}
\usepackage{braket}
\usepackage{hyperref}
\usepackage{titlesec}
\usepackage{balance}
\usepackage{booktabs}
\usepackage{tabularx}
\usepackage{float}
\usepackage{tikz}
\usepackage{fontawesome}
\usepackage[skip=2pt]{caption}
\usepackage{subcaption}
\usepackage{fancyhdr}
\usepackage{float}
\usepackage{placeins}
\usepackage[nameinlink]{cleveref}

\graphicspath{{plots}}
\addto\extrasenglish{

}
\renewcommand{\vec}[1]{\boldsymbol{\mathbf{#1}}} 

\newenvironment{nalign}{
    \begin{equation}
    \begin{aligned}
}{
    \end{aligned}
    \end{equation}
    \ignorespacesafterend
}

\title{Quantum-Assisted Solution Paths for the Capacitated Vehicle Routing Problem}

\fancypagestyle{specialfooter}{%
  \fancyhf{}
  
  \fancyfoot[R]{ \noindent\fbox{%
    \parbox{\textwidth}{%
        {\footnotesize This work has been submitted to the IEEE for possible publication. Copyright may be transferred without notice, after which this version may no longer be accessible.}
        }
    }}
}

\author{
    \IEEEauthorblockN{Lilly Palackal\IEEEauthorrefmark{1}\IEEEauthorrefmark{2}\IEEEauthorrefmark{4}, Benedikt Poggel\IEEEauthorrefmark{3}\IEEEauthorrefmark{4}, Matthias Wulff\IEEEauthorrefmark{1}, Hans Ehm\IEEEauthorrefmark{1}, Jeanette Miriam Lorenz\IEEEauthorrefmark{3}, Christian B.~Mendl\IEEEauthorrefmark{2}}
    \IEEEauthorblockA{\IEEEauthorrefmark{1}Infineon Technologies AG, Munich, Germany}
    \IEEEauthorblockA{\IEEEauthorrefmark{2}Technical University of Munich, Germany}
    \IEEEauthorblockA{\IEEEauthorrefmark{3}Fraunhofer Institute for Cognitive Systems IKS}
     \{lilly.palackal, matthias.wulff, hans.ehm\}@infineon.com \\
     \{benedikt.poggel, jeanette.miriam.lorenz\}@iks.fraunhofer.de,   christian.mendl@tum.de
     
     \IEEEauthorblockA{\IEEEauthorrefmark{4}Authors contributed equally}
}

\date{\today}

\begin{document}
\maketitle

\thispagestyle{empty} 
\pagestyle{plain}
\thispagestyle{specialfooter} 

\begin{abstract}
Many relevant problems in industrial settings result in NP-hard optimization problems, such as the Capacitated Vehicle Routing Problem (CVRP) or its reduced variant, the Travelling Salesperson Problem (TSP). Even with today's most powerful classical algorithms, the CVRP is challenging to solve classically. Quantum computing may offer a way to improve the time to solution, although the question remains open as to whether Noisy Intermediate-Scale Quantum (NISQ) devices can achieve a practical advantage compared to classical heuristics. The most prominent algorithms proposed to solve combinatorial optimization problems in the NISQ era are the Quantum Approximate Optimization Algorithm (QAOA) and the more general Variational Quantum Eigensolver (VQE). However, implementing them in a way that reliably provides high-quality solutions is challenging, even for toy examples. In this work, we discuss decomposition and formulation aspects of the CVRP and propose an application-driven way to measure solution quality. Considering current hardware constraints, we reduce the CVRP to a clustering phase and a set of TSPs. For the TSP, we extensively test both QAOA and VQE and investigate the influence of various hyperparameters, such as the classical optimizer choice and strength of constraint penalization. Results of QAOA are generally of limited quality because the algorithm does not reach the energy threshold for feasible TSP solutions, even when considering various extensions such as recursive and constraint-preserving mixer QAOA. On the other hand, the VQE reaches the energy threshold and shows a better performance. Our work outlines the obstacles to quantum-assisted solutions for real-world optimization problems and proposes perspectives on how to overcome them.
\end{abstract}
\section{Introduction}
Transportation and logistics are becoming increasingly intelligent with the aid of the Internet of Things (IoT), which plays a pivotal role in achieving a smart industry pursuing digitalization and decarbonization. Waste management is an area with high potential for improvement in effectiveness and efficiency. For instance, in large cities, overflowing waste containers are a common problem. Trucks follow fixed routes regardless how full the containers are. To improve these routes, containers can be equipped with sensors providing real-time information on their filling level. Then, optimal routes can be calculated dynamically to only visit full containers and make the most of the truck's capacity. However, efficient planning of vehicle routes is essential to take advantage of these opportunities. The route planning for waste collection translates to a Capacitated Vehicle Routing Problem (CVRP), which is NP-hard. Therefore, industry-sized problems with hundreds to thousands of nodes can only be solved heuristically. 

Quantum computing (QC) offers exciting possibilities to perform certain computational tasks better than classical solvers and may be able to speed up challenging optimization problems like route planning. Powerful algorithms, such as Shor's factorization \cite{shor_1994} or Grover's search \cite{grover_1996} have a proven advantage over their classical counterparts, but require fault-tolerant quantum computers. The current quantum hardware, however, is prone to errors and limited in the number of qubits and connectivity, which sparked interest in variational quantum algorithms for these Noisy Intermediate-Scale Quantum (NISQ) devices~\cite{preskill_quantum_2018}. They use a combination of relatively shallow parameterized quantum circuits and classical optimizers to improve the noise resilience. As a result, these algorithms do not require fully error-corrected qubits and can be used even with relatively few physical qubits. While the capabilities of variational quantum algorithms are not fully understood yet, they have the potential to harness the power of quantum computing even in the current NISQ era. Countless proposals have been put forth, mainly centered around the Quantum Approximate Optimization Algorithm (QAOA)~\cite{farhi_quantum_2014} and the Variational Quantum Eigensolver (VQE)~\cite{Peruzzo_2014}. However, their practical application to industry-relevant optimization problems poses challenges such as finding useful mathematical formulations, setting good hyperparameters of variational algorithms, or measuring the solution quality. For example, a challenge commonly faced when training parameters of a variational quantum algorithm are barren plateaus in the energy landscape~\cite{mcclean_barren_2018}. Also, not all optimization problems might be well-suited for QC. An individual analysis of each application is required.

This work aims to study the application of quantum-enhanced algorithms to an optimization problem motivated by the real world, the CVRP, and to highlight the challenges it poses for QC. We assess the viability of formulation, decomposition, and algorithmic options for NISQ hardware with regards to problem-specific performance measures. To facilitate the numerical experiments, it is necessary to reduce the CVRP to its simpler variant, the Travelling Salesperson Problem (TSP) which arises from the CVRP as a subproblem after an initial clustering step. Many real-world TSP instances can be solved efficiently by classical solvers like Concorde~\cite{concorde_website, Applegate_TSP_2011}, however, a solution to the CVRP remains hard to compute in practice. In this paper, conclusions for quantum-assisted solutions are drawn for both, the TSP as well as the CVRP. We test QAOA and VQE extensively for the TSP and analyze the influence of various hyperparameters like the classical optimizer choice and strength of constraint penalization. We pinpoint what makes TSP hard for QC and work towards separating fundamental problems from the challenge of finding good hyperparameters. Furthermore, we define a performance metric suited to the particular problem which can serve as template for other optimization problems and puts the focus on the solution quality from an application perspective. Obtaining useful results with the QAOA and its variants proves to be difficult even for simple TSPs. We conjecture that the bad results of QAOA are caused by the fact that it does not reach the energy threshold required for feasible TSP solutions that can be found in VQE simulations. Common extensions of QAOA, like recursive QAOA \cite{bravyi_obstacles_2020} or the constraint preserving mixer QAOA \cite{hadfield_AOA_2019} do not yield better results. The VQE, on the other hand, converges better and finds good solutions for all tested instances.

In \autoref{sec: Preliminaries} we introduce the basic concepts in quantum optimization used throughout the paper. In  \autoref{sec: Math. form of use case} we derive QUBO formulations for the CVRP and the subproblems arising from its two-phase solution (Clustering and the TSP) and define an application-specific performance metric. \autoref{sec: quantum solutions to TSP} contains the main results of this paper, where the TSP is solved with QAOA, VQE and several extensions. We discuss the results in \autoref{sec: discussion} and suggest future research directions in \autoref{sec: outlook}.
\section{Preliminaries}\label{sec: Preliminaries}

\subsection{Quadratic Unconstrained Binary Optimization}
Many discrete optimization problems can be formulated in terms of a Quadratic Unconstrained Binary Optimization (QUBO) problem. It is a convenient input format for quantum optimization algorithms, stated as minimization problem $\min_x f(x)$ with binary decision variables $x\in \{0,1\}^n$, and a quadratic cost function $f(x) = x^T Q x$ for a square matrix $Q \in \mathbb{R}^{n\times n}$, the QUBO matrix. Thus, a QUBO problem is defined by

\begin{equation}
    \min_{x\in \{0,1\}^n} x^T Q x.
\end{equation}
Any QUBO can be mapped directly to an Ising model and vice versa~\cite{barahona_experiments_1989} which is why it is extensively used in QC. This transformation is sketched in \autoref{subsection: The TSP} for the TSP.

\subsection{Variational Quantum Algorithms} \label{subsection:VQA}

Variational quantum computing \cite{cerezo_variational_2021} is a proposal to make use of imperfect NISQ era devices. It combines shallow parameterized quantum circuits with classical optimizers, aiming to find a combination of parameters such that the circuit prepares a quantum state representing the solution of a given problem. The main application areas are simulation (e.g. in quantum chemistry), machine learning and mathematical optimization. The latter starts with expressing the cost function $C(\vec{x})$ as a quantum expectation value:

\begin{equation}
    C(\vec{x}) = \braket{\psi(\vec{x}) | H_P| \psi(\vec{x})}
    \label{eq:cost_function_as_braket}
\end{equation}
The state $\ket{\psi(\vec{x}})$ is prepared by a quantum circuit (the \emph{ansatz}) depending on parameters $\vec{\theta}$. Mathematically, this changes the optimization variables to $\vec{\theta}$.

\begin{nalign}
    \label{eq:vqa_ansatz}
    \ket{\psi (\vec{\theta})} &= U(\vec{\theta}) \ket{+} \\
    C(\vec{\theta}) &= \braket{+|U^\dagger(\vec{\theta}) H_P U(\vec{\theta}) | +}
\end{nalign}
Once the ansatz is chosen, a classical optimizer tries to find the optimal parameters $\theta$ that minimize $C(\vec{\theta})$ by repeatedly evaluating the expectation value in \eqref{eq:vqa_ansatz}.

QAOA is a particular ansatz, originally proposed in \cite{farhi_quantum_2014}, that can be regarded as the trotterized version of adiabatic quantum computing. Given the optimization problem encoded into an Ising Hamiltonian $H_P$, the ansatz is then given by

\begin{equation}
    U(\vec{\beta}, \vec{\gamma}) = \prod_{k=1}^p e^{-i\beta_k H_M} e^{-i\gamma_k H_P}.
    \label{eq:qaoa_ansatz}
\end{equation}
The number of alternating layers $p$ is commonly called \emph{depth}. $H_M$ is the \emph{mixer Hamiltonian}, corresponding to the driver in adiabatic quantum computing, Its standard choice is $H_M = \sum_j \sigma_j^x$. 

Starting from this simple variational protocol, numerous improvements have been proposed such as warm-start \cite{egger_warm-starting_2021} and recursive variants \cite{bravyi_obstacles_2020}.

VQE~\cite{Peruzzo_2014} is a more general variational quantum algorithm allowing for an arbitrary ansatz. Its advantage over QAOA for optimization problems lies in the availability of hardware-efficient ans\"atze that are designed specifically to respect the native gate set and connectivity of the quantum hardware. The ansatz in \eqref{eq:vqa_ansatz} is

\begin{nalign}
    U(\vec{\theta}) &= \prod_{k=1}^{p+c} U_k(\theta), \\
    \label{eq:vqe_ansatz}
\end{nalign}
where the circuit may contain $c$ non-parameterized gates and $p$ parameterized ones of the form $e^{-i\theta H_k}$. Usually, $p$ is much larger compared to QAOA, but each operator $U_k$ typically only involves one or two qubits.
\section{Mathematical Formulation of the Use Case}\label{sec: Math. form of use case}

\begin{figure}[H]
    \centering
    \begin{tikzpicture}[node distance={2cm}, thick, main/.style = {draw, circle}]
    \node[scale=2] (T) at (3,-4.5) {\faTruck};
    \node[scale=2] (c) [below of=T, node distance=3mm] {\tiny capacity $=10$};
    \node[main, fill = blue!20] (0) {D};
    \node[main, fill = blue!20] (1) [below left of=0, node distance=10mm] {$1$};
    \node[main, fill = blue!20] (3) [below left of=1] {$3$};
    \node[main, fill = blue!20] (2) [below right of=3] {$2$};
    \node[main, fill = blue!20] (4) [below left of=2] {$2$};
    \node[main, fill = blue!20] (5) [right of=2] {$2$};
    \node[main] (6) [right of=0] {$5$};
    \node[main] (7) [above right of=6] {$2$};
    \node[main] (8) [above left of=7] {$1$};
    \node[main] (9) [above of=0] {$5$};
    \node[main] (10) [left of=9] {$5$};
    \draw[->,style={blue,draw=blue}] (0) to node[midway, above left] {$10$} (1);
    \draw[->,style={blue,draw=blue}] (1) to node[midway, above left] {$21$} (3);
    \draw[->,style={blue,draw=blue}] (3) to node[midway, above right] {$33$} (2);
    \draw[->,style={blue,draw=blue}] (2) to node[midway, above left] {$42$} (4);
    \draw[->,style={blue,draw=blue}] (4) to node[midway, below right] {$47$} (5);
    \draw[->,style={blue,draw=blue}] (5) to node[midway, above right] {$38$} (0);
    \draw[->] (0) to node[midway, above right] {$31$} (6);
    \draw[->] (6) to node[midway, below right] {$45$} (7);
    \draw[->] (7) to node[midway, right] {$20$} (8);
    \draw[->] (8) to node[midway, left] {$30$} (0);
    \draw[->] (0) to node[midway, left] {$42$} (9);
    \draw[->] (9) to node[midway, above] {$15$} (10);
    \draw[->] (10) to node[midway, left] {$30$} (0);
\end{tikzpicture}
    \caption{An example instance of the capacitated vehicle routing problem and its optimal path. The number in each node corresponds to its demand, the edge weights their distances, and ``D'' marks the depot. The blue cycle will be used as 6-, 5- and 4-node TSP instances for the numerical experiments.}
    \label{fig:CVRP_instance}
\end{figure}
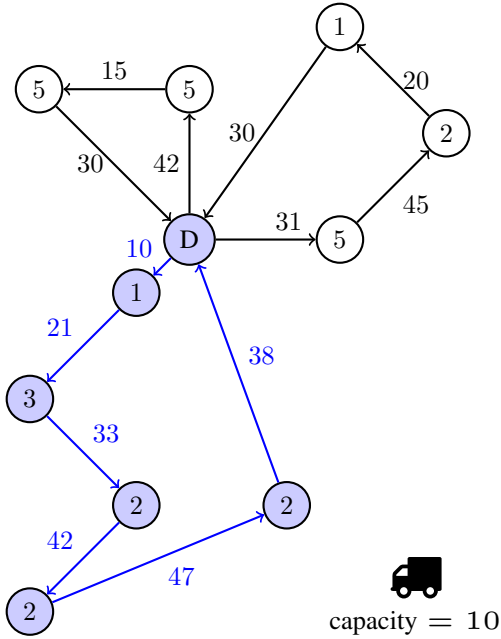

The optimization problem motivated by the waste management logistics is modeled as a CVRP. Formally, the CVRP is defined on a graph with $n$ nodes and weighted directed edges. Each node $i$ represents a customer with demand $d_i$, where node $0$ represents a depot. The edge weights $w_{ij}$ represent distances for travelling from node $i$ to $j$. We aim to find the shortest route for a vehicle with capacity $C$ visiting each customer to satisfy their demand. When the vehicle visits a customer node $i$, its current capacity is decreased by $d_i$. Upon returning to the depot $0$, the capacity is filled up to $C$ again. 
If $\sum_{i=1}^n d_i \leq C$, the CVRP reduces to searching for a tour among all vertices with least total edge weight, known as the travelling salesperson problem (TSP). It is described in more detail in \autoref{subsection: The TSP}.
However, in the case $\sum_{i=1}^n d_i > C$, the optimal tour will split up into cycles intersecting at the depot node as shown in \autoref{fig:CVRP_instance}. The TSP and by extension the harder CVRP are proven to be NP-hard \cite{Karp_TSPisNPhard_1972}. 

\subsection{QUBO formulation of the CVRP}\label{subsection: QUBO for CVRP}

In recent years, several QUBO formulations have been stated for CVRP variants. Irie et al.~\cite{Irie_CVRPasQUBO_timebased_2019} introduce a time-table based CVRP formulation defining decision variables which indicate whether a vehicle arrives at a location within a certain time interval. Harikrishnakumar et al.~\cite{Harikrishnakumar_MDCVRP_2020} allow for CVRPs with multiple depot nodes.
The CVRP formulated in \cite{Irie_CVRPasQUBO_timebased_2019} and \cite{Harikrishnakumar_MDCVRP_2020} considers one vehicle visiting all customers which is able to return to the depot to fill up its capacity. In this work, we consider several vehicles with capacity $C \in \mathbb{N}$ being sent to visit some customers and return to the depot without the possibility of recharging at the depot. We optimize the total length of all paths such that $\text{CO}_\text{2}$ emissions are minimized. The CVRP with multiple vehicles is also considered by Feld et al.~\cite{Feld_2019}. However, our formulation needs fewer terms and requires slightly less qubits. For this, let the binary variable $x_{v,t}^k$ mark whether vehicle $k$ visits customer $v$ at time step $t$. For an ideal tour, we need $K^*$ vehicles to serve all customers. In general, we do not know $K^*$ a priori, thus, we suggest the number of vehicles to be set to $K=\left\lceil \frac{\sum_{v\in V} d_v}{C}\right\rceil$. Similarly, $T^*$ is unknown and is set as $T= n$ for the worst case  of one vehicle visiting all costumers. 
The objective function to be minimized is the total distance given by 

\begin{equation}
    H_{\text{obj}} = \sum_{v,v',t, k} w_{vv'} \cdot x_{v, t}^k \cdot x_{v', t+1}^k.
\end{equation}
The three constraints $(C1)$ visiting each customer exactly once, $(C2)$ visiting exactly one customer or the depot for each time step and each car, and $(C3)$ respecting the capacity of each vehicle, can be formulated as
\begin{nalign}
(C1)& \quad H_{C_1}=\sum_{v=1}^{n}\left(1 - \sum_{k=1}^K \sum_{t=1}^T x_{v,t}^k\right)^2\stackrel{!}{=}0,\\
    (C2)& \quad H_{C_2}= \sum_{k=1}^K \sum_{t=1}^T \left(1 - \sum_{v=0}^{n} x_{v,t}^k\right)^2 \stackrel{!}{=}0,\\
    (C3)& \quad \sum_{t=1}^T\sum_{v=1}^n x_{v,t}^k \cdot d_v \leq C \quad \forall k\in K.
\end{nalign}
In order to ensure that every vehicle starts and ends at the depot we set $x^k_{0,1}=x^k_{0,T}=1$ for each $k$. In general, we might have chosen a larger number of time steps than necessary for the optimal solution, i.e. $T> T^*$. Whenever a time step is not needed, the constraint $(C2)$ will force $x^k_{0,t}=1$ for all time steps $t$ corresponding to the car $k$ staying at the depot. Note that this does not contradict constraint $(C1)$ as the sum over the nodes starts at $v=1$.
Next, we transform the inequality in $(C3)$ into an equation by introducing slack variables $y^k_b\in \{0,1\}$ which represent the sum of demands of customers visited by vehicle $k$. With a logarithmic encoding, this requires $K\cdot \left \lceil \log_2 (C+1) \right \rceil$ additional qubits resulting in the QUBO term

\begin{nalign}
    H_{C_3}=&\sum_{k=1}^K\left[ \left(\sum_{t=1}^T\sum_{v=1}^n x_{v,t}^k \cdot d_v\right)\right.\\
   &\left.  +\left(\sum_{b=0}^{\lceil\log (C+1) \rceil -1}2^b \cdot y^k_b\right) -C \right]^2.
\end{nalign}
A complete QUBO formulation for the CVRP is thus given by

\begin{equation}
    H_{\text{CVRP}} = H_{\text{obj}} + P_1 H_{C_1}+ P_2 H_{C_2}+ P_3 H_{C_3}.
\end{equation}
This QUBO requires fewer resources (terms and qubits) than the formulation derived in \cite{Feld_2019}. Instead of $K\cdot C$ slack bits, we use $K\cdot \left \lceil \log_2 (C+1) \right \rceil$.
However, a large number of decision variables, $n\cdot T \cdot K $, remains. For NISQ devices, it is thus necessary to consider a heuristic two-phase approach splitting the CVRP into a Clustering Phase (CP) and a TSP \cite{Laporte_heuristicsCVRP_2002}. This method drastically reduces the number of qubits at the cost of possibly losing optimality. The heuristic works in two steps: the CP aims to group customers such that the demands within a group do not exceed the truck's capacity and the groups stay close together. Then, it remains to solve the TSP for each cluster.

\subsection{The Clustering Phase} \label{subsec: QUBO of Clustering}

The CP is modelled as a multiple Knapsack problem, where each customer should be assigned to a knapsack such that the sum of the customers' demands does not overstep the knapsack's capacity $C$, which is equal to the truck's capacity from the CVRP. Additionally, the sum of distances within each Knapsack is minimized. A QUBO formulation to the CP is given in \cite{Feld_2019}. We slightly change this formulation by introducing a binary encoding for the slack variables.

\begin{align}
\begin{split}
    P_1 \sum_{k=1}^K\left[ \left(\sum_{v=1}^n x_{v}^k \cdot d_v\right) +\left(\sum_{b=0}^{\lceil\log (C+1) \rceil-1}2^b \cdot y_{b}^k\right) -C\right]^2 \\
    + P_2 \sum_{v=1}^n \left( 1-\sum_{k=1}^K x_v^k \right)^2 + P_3 \sum_{k=1}^K \left( \sum_{u,v \in V} D_{uv} x_u^k x_v^k\right), \label{eq: QUBO for Clustering}
\end{split}
\end{align}
where $x_v^k \coloneqq \sum_{t=1}^{T} x_{v, t}^k$ and $x_v^k =1$, if customer $v$ is packed in Knapsack $k$. The slack variables $y_b^k$ represent the filling level of each knapsack and $K=\left\lceil \frac{\sum_{v\in V} d_v}{C}\right\rceil$ is the number of knapsacks as defined in \autoref{subsection: QUBO for CVRP}. A penalty factor $P_1$ ensures that each knapsack's filling level takes a value between $0$ and the capacity $C$. Analogously, $P_2$ requires each customer to be packed into exactly one knapsack. 
The last term in \eqref{eq: QUBO for Clustering} is the sum of all edge weights within a cluster. As the last term can easily dominate the other summands, especially for large $C$, we aim to normalize it. With an approximate number $\frac{n}{K}$ of customers per cluster, a route within a cluster visiting each node once contains $\frac{n}{K}-1$ edges. Therefore, we suggest to choose the penalty factor $P_3= p \cdot \frac{K}{n}$, where $p$ is a constant. With this normalization, the summand approximately represents the length of a route in a given cluster.

As described in \autoref{subsec: QUBO of Clustering}, the CP is similar to a multiple knapsack problem extended by the minimization of the sum of all distances within a knapsack. In~\cite{Awasthi_multiKnapsack_2023}, several quantum approaches to the multiple Knapsack problem are evaluated demonstrating a rather bad performance. This aligns with the performance benchmarks we derived for the CP of the CVRP when running QAOA and VQE. Therefore, we do not further investigate the performance of these algorithms for the CP in this work.

\subsection{The Travelling Salesperson Problem}\label{subsection: The TSP}
The TSP can be stated as follows: Given a complete, undirected and weighted graph, find the shortest cycle that contains all vertices. In this picture, cities are modeled as vertices and the distances between them as edge weights of the graph. A minimal complete description of a problem instance is therefore given by its (symmetric) \emph{adjacency matrix} $D$.

The direct encoding into a QUBO is given in \cite{lucas_ising_2014}. It requires $(n-1)^2$ binary decision variables $x_{ij}$, where $i$ labels the cities and $j$ the position they appear in the path. The reduction from $n^2$ to $(n-1)^2$ variables comes with the elimination of the cyclic permutation symmetry of the TSP paths. For simplicity, we present the non-reduced $n^2$-variable cost function:

\begin{align}
\begin{split}
    C(\{x_{ij}\}) = s \left[ \sum_{i,i',j = 1}^n D_{ii'} x_{ij} x_{i'(j+1)}\right.\\
     \left.+ P \sum_{i=1}^n \left(1 - \sum_{j=1}^n x_{ij}\right)^2\right. \\
     \left. + P \sum_{j=1}^n \left(1 -  \sum_{i=1}^n x_{ij}\right)^2 \right].
     \end{split}
\end{align}
Here, $s$ is an overall scaling factor that does not change the optimum, but may be used to improve numerical stability and convergence. $P$ scales the penalty associated to the two constraints of (C1) visiting each city exactly once and (C2) asserting each position exactly once. In the first sum, the actual TSP path length, $n+1 \equiv 1$ is assumed. The reduction to $(n-1)^2$ variables is straightforward by setting $x_{11}=1$ and $x_{1i} = x_{i1} = 0 \; \forall i \neq 1$. Since all variables are binary, $x_{ij}^2 = x_{ij}$ and the corresponding QUBO matrix element $Q_{ijkl}$, the interaction between $x_{ij}$ and $x_{kl}$, is given by

\begin{align}\label{eq: QUBO tensor}
\begin{split}
    Q_{ijkl} = -2sP\delta_{ik}\delta_{jl} + sP [\delta_{ik}(1-\delta_{jl}) + \delta_{jl}(1-\delta_{ik})] \\
    + sD_{ik} [\delta_{j(l+1)} + \delta_{j(l-1)}].
\end{split}
\end{align}
The first addend describes the diagonal terms, the second the penalty for violating the constraints, and the third the TSP path length. A constant contribution $o_Q \coloneqq 2nsP$ (the QUBO offset) has been omitted here such that the QUBO problem reads

\begin{nalign}
    &\textrm{Minimize} \quad \sum_{i,j,k,l} x_{ij} Q_{ijkl} x_{kl} \\
    &\textrm{for} \quad x_{ij} \in \{0,1\} \forall i,j = 1,\dots, n .
\end{nalign}
One can now pass to an Ising formulation with variables $z_{ij} \in \{-1,1\}$ via $z_{ij} \coloneqq 2 x_{ij} - 1$, dropping another offset $o_I = \sum_{i,j,k,l} Q_{ijkl} - 2sn^2P$ along the way (the latter contribution arises because $z_{ij}^2 = 1$). Promoting the variables to Pauli-Z operators $\sigma_z^{ij}$ then yields the Ising Hamiltonian

\begin{equation}
    H_P = \sum_{(i,j) \neq (k,l)} \frac{Q_{ijkl}}{4}\sigma_z^{ij} \sigma_z^{kl} + \sum_{i,j,k,l} \frac{Q_{ijkl}}{2} \sigma_z^{kl} .
\end{equation}
The open questions are now how to determine $s$ and $P$ in~\eqref{eq: QUBO tensor}, and how to judge the quality of a state in the presence of potentially infeasible states. For the sake of completeness, the TSP can be cast into different formulations by representing a path as a permutation of $(1,\dots, n)$ and then encoding these integer variables into binaries. The formulation used here is equivalent to one-hot encoding in this protocol.

\subsubsection{A performance metric}\label{subsubsec: Performance metric}
In order to compare the performance of quantum algorithms on the TSP, a performance metric akin to the MaxCut approximation ratio is needed. For the TSP, infeasible solutions do not correspond to routes and are thus useless. A desirable metric should yield information about the overlap of an outcome $\ket{\psi}$ with the subspace of feasible solutions. For this reason, we propose an alternate measure consisting of a pair of values. First, we would like to know $\sum_{\ket{\text{feasible}}} |\braket{\text{feasible}|\psi}|^2$. Since for the TSP the number of feasible solutions grows as a factorial of the problem size, calculating this quantity will quickly become costly when scaling the problem size. Instead, we suggest to consider the \emph{feasibility ratio}

\begin{equation}
    m_{\text{feas}}\coloneqq \frac{\textrm{\# feasible shots}}{\textrm{\#shots}}.
\end{equation}
Similarly, when searching for an optimal (and not only feasible) state, we are interested in the overlap of optimal states with feasible ones  $\sum_{\ket{\text{optimal}}} |\braket{\text{optimal}|\psi_{\text{feasible}}}|^2$, where $\ket{\psi_{\text{feasible}}}$ is the projection of $\ket{\psi}$ onto feasible states $\sum_{\ket{\text{feasible}}} \ket{\text{feasible}} \braket{\text{feasible}|\psi}$. However, this sum should be weighted with the TSP path length to measure the closeness to the optimal solution. Ultimately, optimality can then be measured by the \emph{TSP length ratio}
\begin{equation}\label{eq: perf. metric TSP length}
    m_{\text{len}}\coloneqq \frac{\textrm{optimal TSP path length}}{\textrm{averaged TSP path length for feas. shots}}.
\end{equation}
Since the optimal TSP path length is, in general, unknown, the nominator can be replaced with the best known classical solution.
Finally, an informative measure of the performance of a quantum algorithm for the TSP is given by the tuple

\begin{equation}
    \left(m_{\text{len}}, m_{\text{feas}} \right).
\end{equation}
These indicators are well suited for assessing the quality of a TSP solution even in the absence of a working amplification of the solution state. Unlike the energy, they depend on the penalty only through the optimization process and can thus be used to compare different penalty values and selection strategies. Naturally, the ideal value is $(1,1)$ and the equal superposition of all basis states corresponds to $(c, \frac{(n-1)!}{2^{(n-1)^2}})$, where $c=\frac{\textrm{optimal TSP path length}}{\textrm{averaged TSP path length for } \textbf{all} \textrm{ feas. states}}$ depends on the given instance. This quickly goes to $(c,0)$ with increasing $n$. Note that the TSP length ratio is bounded from below by ratio corresponding to the longest path $l$ ($l< m_{\text{len}}<1$).
Further, there is a potentially high variance in determining the ratios in a shot-based way, especially when the algorithm fails to amplify the solution states strongly enough and the circuit therefore samples from a very flat distribution.

This performance measure can analogously be defined for the CVRP by simply exchanging the TSP length by the CVRP length in \eqref{eq: perf. metric TSP length}.

\section{Quantum-Assisted Solution Paths for the TSP}\label{sec: quantum solutions to TSP}

We now turn to the solution of the TSP. The algorithms are tested on instances with 4, 5 and 6 nodes derived from the blue part in the CVRP (\autoref{fig:CVRP_instance}). These instances are encoded into 9-, 16- and 25-qubit Hamiltonians such that the resulting algorithms can be run with a statevector simulator. We extensively test QAOA and VQE as well as the hyperparameters. Beyond the scaling $s$ and penalty factor $P$ as described in \autoref{subsection: The TSP}, those include algorithm-specific options such as the optimizer or the exact ansatz. 

\begin{figure}
    \centering
    \includegraphics[width=\linewidth]{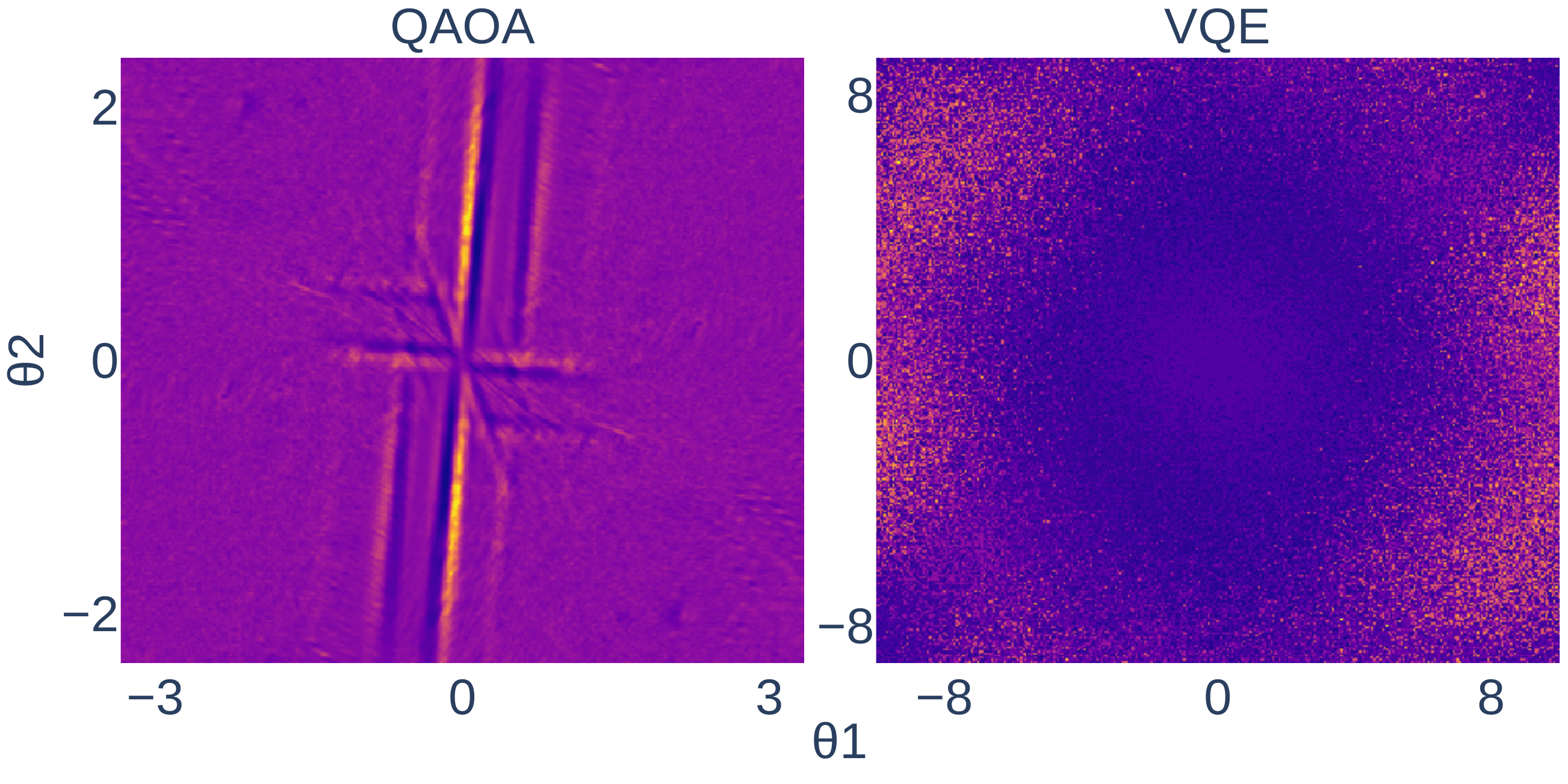} 
    \caption{The loss landscapes for QAOA (left) and VQE (right) differ drastically. Both are plotted in a plane randomly oriented in the parameter space of 10 (QAOA) and 27 (VQE) dimensions. The overall structure of the loss landscape is much less complicated for VQE, but for large parameter values, high fluctuations come into play. QAOA clearly has a large area with near-vanishing gradients with maxima and minima close by each other.}
    \label{fig:losslandscapes}
\end{figure}

VQE and QAOA behave differently due to their loss landscapes. \autoref{fig:losslandscapes} compares a 9-qubit QAOA ansatz with depth 5 and a hardware-efficient 9-qubit VQE with 27 parameters. It shows a plane parameterized by $\theta_1$ and $\theta_2$ that is randomly oriented in the high-dimensional parameter space. VQE exhibits a much coarser structure, although high-frequency components appear for larger parameter values. For QAOA, maxima and minima lie closely together, surrounded by areas with small gradients. Taken together, it seems like there is a tradeoff between the number of dimensions of the parameter landscape and its complexity for single parameters. 

\subsection{QAOA}
\label{sec:qaoa}

The QAOA ansatz is fully determined by the problem Hamiltonian $H_p$ depth $p$. The algorithm furthermore needs a classical optimizer and an initial point. The complete selection of hyperparameters of the algorithm and Hamiltonian formulation is given in \autoref{tab:hyperparams_TSP}. Since finding good parameter selection strategies is an open research question, especially combined with an application-specific performance metric, we turn to discussing the tuning options for these parameters.

\begin{table}
\caption{Hyperparameter choices for setting up QAOA for the TSP}
\label{tab:hyperparams_TSP}
\begin{center}
\begin{tabularx}{\linewidth}{lcll}
    \toprule
    \multicolumn{2}{c}{Hyperparameter} & Options & Policy \\
    Name & Symbol & & \\
    \midrule
    Penalty Factor & $P$ & $P > P_{min} $ & \autoref{subsection:penalty_factor} \\
    Scaling Factor & $s$ & $s > 0$ & \autoref{subsection:scaling_factor} \\
    QAOA depth & $p$ & $p\in \mathbb{N}$ & fixed $p=5$ \\
    Optimizer & & options provided by Qiskit & Testing \\
    Initialization & & random, linear & Testing \\
    \bottomrule
\end{tabularx}
\end{center}
\end{table}

\subsubsection{Scaling Factor Selection} \label{subsection:scaling_factor}
The scaling factor $s$ does not influence the location of the minimum. However, adjusting $s$ might help the optimizer since it treats the $\beta$ and $\gamma$ parameters on equal grounds~\cite{brandhofer_benchmarking_2022}. It is therefore beneficial to equalize their impact on the ansatz. Ref.~\cite{brandhofer_benchmarking_2022} tries to achieve this by scaling $H_P$ such that its spectral width aligns with the mixer $H_M$. However, the approximate period of the $e^{-i\gamma H_P}$ operator depends on the spacing of the eigenvalues. Therefore, we propose the \emph{ground state gap} strategy instead: equalize the gap between the ground state and lowest excited state. The smallest excitation of the $\sigma_x$ mixer corresponds to a single spin flip with an energy gap $\Delta_M = 2$. For small systems, the ground state gap has a high variance and leads to larger $s$ overall. With increasing system size, both strategies approach each other.

Although the scaling factor selection is not the most pressing issue, it should not be neglected as it directly influences the unit in which the distances between the TSP nodes are to be measured. For toy examples, both the spectral width and ground state gap strategies roughly amount to normalizing the distances such that they are of $\mathcal{O}(1)$.

\subsubsection{Penalty Factor Selection} \label{subsection:penalty_factor}

The penalty factor needs to ensure that the lowest energy state is feasible and thus optimal. It is therefore bounded from below by a $P_{min}$ and unbounded from above. From a practical standpoint, a certain separation between the optimal state and infeasible ones is desirable, but the penalty should not be chosen too high to avoid very rugged loss landscapes getting in the way of the optimization process. The test instances are classically solvable and $P_{min}$ can be calculated, but it is unknown for unsolved real-world instances. We therefore adopt the following protocol: Determine $P_{min}$ for a number of classically accessible instances of increasing size, then derive a scaling relation to approximate $P_{min}$ for larger instances. The actual penalty should then be chosen slightly larger to achieve good penalization (e.g., $1.2\cdot P_{min}$). Computing $P_{min}$ for TSP sizes 4 through 9 and 1000 instances each shows that it does not scale with system size, but remains stable around $P_{min}=57 \pm 11$ with fewer than $0.1\%$ problematic outliers above the $3\sigma$ interval. Luckily, a sanity check for the feasibility of obtained solutions is easy.

\subsubsection{Initialization}

The QAOA loss landscapes exhibit large areas of barren plateaus~\cite{mcclean_barren_2018} and sections riddled with local minima. Therefore, the initial point in parameter space plays an important role in the performance of QAOA. For the scope of this paper, we consider two relatively simple strategies, uniformly random initialization and a linear initialization. The latter is inspired by the analogy between QAOA and quantum annealing. The $\beta$ parameters decrease linearly with each layer whereas the $\gamma$ increase linearly in order to mimic the annealing time evolution. We observe no statistically significant difference between the two strategies and, for this reason, stick with random initialization. In principle, there are more sophisticated strategies for the initial point selection~\cite{amosy_iterative-free_2022}.

\subsubsection{QAOA Depth}

The depth $p$ of QAOA determines the number of alternating layers applied to construct the ansatz. As $p$ increases, the approximation of the ground state of the problem Hamiltonian is expected to improve~\cite{farhi_quantum_2014}. Due to decoherence, NISQ devices are limited to relatively shallow circuits. Following a reverse causal cone argument~\cite{streif_training_2019}, a depth of $p=5$ should be sufficient to produce reasonable solutions.

\subsubsection{Simulating QAOA with Various Optimizers and Hyperparameters}

Together with the hyperparameters shown above, the optimizers accessible through Qiskit are tested in statevector simulation for the 4-node TSP instance (for data see \Cref{app:qaoa}). Throughout all tests, the achieved feasibility ratio is low, and the variational procedure does not manage to amplify a solution state sufficiently, or reliably produce a feasible state. This is true even for the smallest example instance of four cities. The optimizers still differ in performance as seen by the final energies, with NFT and UMDA reaching the best states. Among the other optimizers, some fail to converge with the allocated iteration budget (Nelder-Mead, SPSA) whereas others finish quickly without any meaningful optimization effort (SLSQP). While still far from any useful solution, the Powell optimizer shows the best combination of the number of quantum circuit executions and energy obtained. 

In light of these results, tests with real hardware do not seem useful. Even switching from the exact statevector simulation method to approximate ones resulted in a severe decrease in solution quality, most likely due to the very flat distribution of output states. Furthermore, real hardware would be much more restricted in the number of circuit executions. Overall, it is doubtful that the performance gap can be closed by mere hyperparameter tuning. Our experiments with various scaling factors, penalties, depth and initial points support this claim. 

\subsection{Warm-Start QAOA}
\label{sec:ws-qaoa}

A possible improvement to the standard QAOA ansatz, referred to as warm-start QAOA (WS-QAOA), was proposed by Egger, Mareček and Woerner~\cite{egger_warm-starting_2021}. Instead of preparing a uniform superposition state $\ket{+}^{\otimes n}$, a classical solver obtains a solution to the relaxed continuous problem, $\Tilde{x}\in [0,1]^n$ that, once encoded, serves as initial state 
\begin{equation}
    \ket{\widetilde{\psi}}= \bigotimes_{i=0}^{n-1} R_Y(\theta_i)\ket{0},
\end{equation}
where the angles are defined by the solution $\Tilde{x}$ by $\theta_i = 2\arcsin{\sqrt{\Tilde{x_i}}}$.
Further, a new mixer operator is defined that depends on the continuous solution $\Tilde{x}$ as well. The solution to the relaxation of the QUBO problem of the 4-node TSP instance turns out to be integer. Thus the initial state is a computational basis state corresponding to the optimal solution and a search with QAOA is not necessary. For this reason, we cannot derive any insights on the performance of WS-QAOA based on the present instance.

\subsection{Recursive QAOA}
\label{sec:rQAOA}

Recursive QAOA (rQAOA)~\cite{bravyi_obstacles_2020} is a potential extension of the QAOA protocol aimed at improving its performance. The idea is to use QAOA to find correlated qubits that can then be merged to reduce the problem size until it can be solved classically. The strongest correlation among qubit pairs is effectively rounded to $\pm 1$, replacing both qubits with a single one while modifying the cost function in accordance with the sign of the correlation. However, the success of this reduction directly depends on the accuracy of the rounding. The low-quality QAOA output achieved for the TSP does certainly not suffice to produce meaningful results. For the MaxCut problem (the standard test problem for QAOA), rQAOA has shown promising results. However, when applied to the TSP, rQAOA encounters two additional obstacles. First, the reduction process can be very fragile with approximate solutions, and even a single error can render the entire rQAOA output meaningless. For example, one faulty correlation that results in setting two qubits associated to the same node as equal results in the whole rQAOA never exploring feasible states in the future. MaxCut has no infeasible states and thus does not suffer from this problem. Second, the one-hot encoding together with the inherent inversion symmetry of the TSP paths significantly reduces the amount of correlation in the qubit pairs. To illustrate, the 4-node TSP has two optimal routes, (0,1,3,2) and (0,2,3,1), corresponding to the bitstrings 100001010 and 010001100, respectively. Any algorithm that only distinguishes states by their Ising energy cannot tell them apart. A perfect solution is therefore a mixture between these states. However, only half of the qubit pairs are correlated. For the two equivalent solution strings of a MaxCut problem in contrast, all qubit pairs are correlated. In the light of these theoretical and practical difficulties, applying rQAOA to the TSP problem does not seem sensible.

\subsection{Constraint-Preserving Mixer QAOA}

The Constraint-Preserving Mixer QAOA, often referred to as Quantum Alternating Operator Ansatz (AOA) \cite{hadfield_AOA_2019}, aims to restrict the search space to the subspace containing the feasible solutions. Consider a binary optimization problem $\min f(x)$ for $x \in F \subset \{0,1\}^n$. Define a Hamiltonian $H_f$ to act as $f$ on computational basis states, i.e. $H_f\ket{x}= f(x)\ket{x}$. Then the Alternating Operator Ansatz (AOA) is defined by two families of parameterized operators,

\begin{itemize}
    \item phase separation operators $U_P(\beta)$ defined by the objective function $f$, e.g. $U_P(\beta)= e^{-i\beta H_f}$, and
    \item mixing operators $U_M(\gamma)$ defined by the constraints that preserve the feasible subspace $F$ and allow its full exploration.
\end{itemize}
The quantum circuit of the AOA is given by the alternating application of the two operators for $p$ layers on an easily prepared feasible initial state $\ket{s}$ ($s\in F$). The parameters are then optimized classically as in QAOA.

In~\cite{hadfield_AOA_2019}, constructions for mixer operators for several applications are given, including the TSP. Let
\begin{align}
    S^+ &= \ket{1}\bra{0}, \quad S^- = \ket{0}\bra{1},\\
    \notag\\
    H_{\text{M}, i, \{u,v\}}&= S_{u,i}^+ S_{v,i+1}^+ S_{u,i+1}^- S_{v,i}^- + S_{u,i}^- S_{v,i+1}^- S_{u,i+1}^+ S_{v,i}^+,\\
    H_\text{M} &= \sum_{i=1}^n \sum_{\{u,v\}\in \binom{[n]}{2}} H_{\text{M}, i, \{u,v\}}.
\end{align}
Then a mixer operator for the TSP is given by 
\begin{equation}
    U_\text{M}(\beta)= e^{i \beta H_\text{M}}.
\end{equation}
For a 4-node TSP instance, this mixer Hamiltonian consists of 1100 terms. In our simulations, states remain feasible throughout the search, however, the optimal solution was not found.

\subsection{Variational Quantum Eigensolver}
\label{sec:vqe}

For VQE, we test a simple ansatz comprising a layer of single-qubit rotation gates (X and Z) followed by entangling controlled-X rotation gates in nearest-neighbor connectivity on a chain topology. To ensure a fair comparison, we adjust the maximum allowed number of iterations to the optimizer to avoid exceeding the timeout of the IBM statevector simulator. All TSP instances were formulated with a penalty of $P=100$ for the optimizer comparison. To facilitate the comparison with the QAOA runs, we chose the scaling factor with the ground state gap strategy outlined in \autoref{subsection:scaling_factor}. However, since the optimizers are scale-invariant and the problem Hamiltonian does not enter the ansatz, the scaling factor is irrelevant for VQE.

\subsubsection{Optimizers}
\label{sec:vqe_opt}

\begin{figure*}[h]
    \centering
    \includegraphics[width=0.9\linewidth]{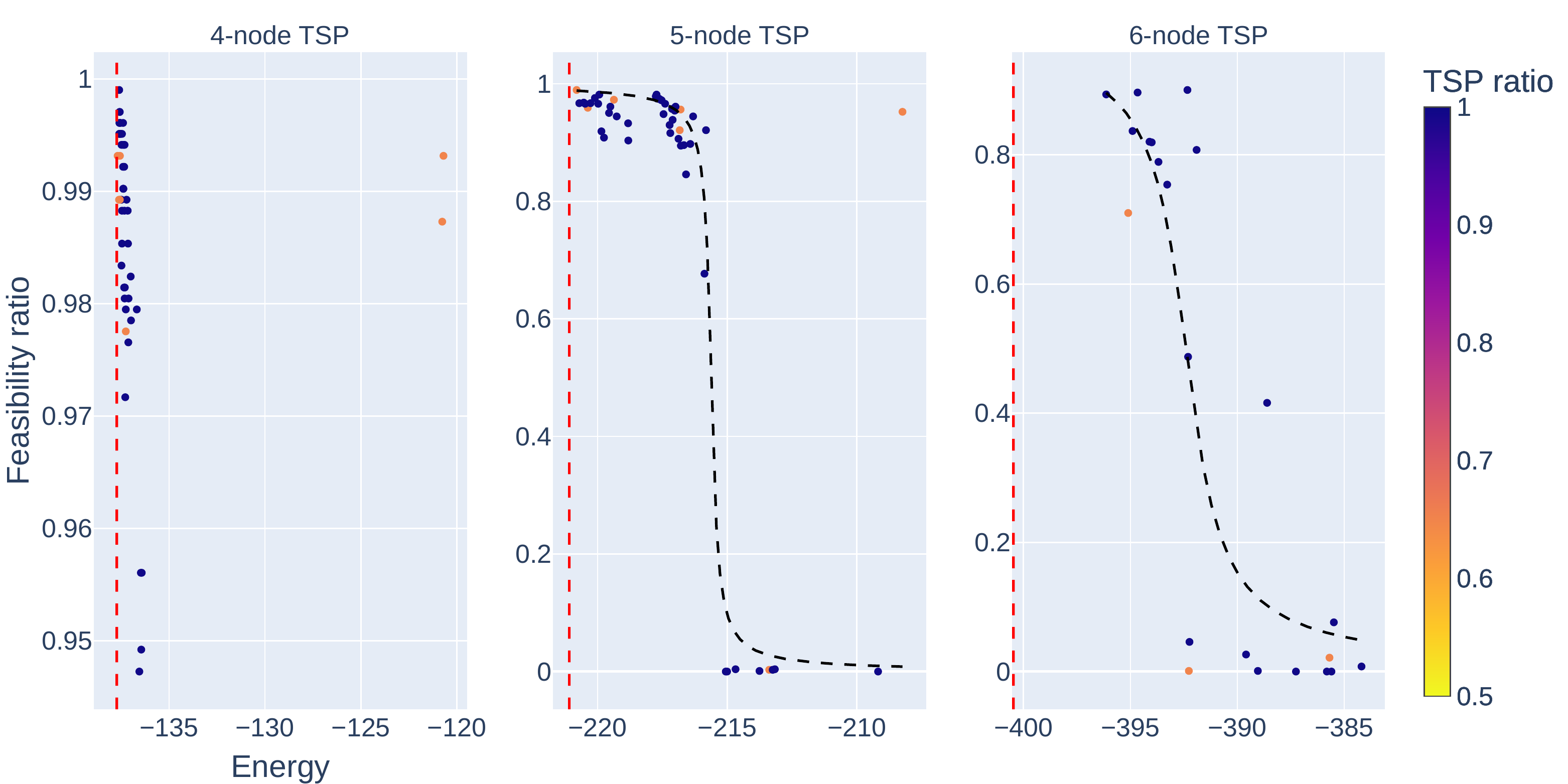}
    \caption{The relation of the feasibility ratio to the Ising energy of VQE with the Powell optimizer on a TSP with 4 to 6 nodes. For the 5- and 6-node TSP, it becomes clear that the feasibility undergoes drastic changes at a transition point in energy, illustrated with an arctan fit. The data was obtained with a QASM Simulator on the IBM cloud. 50 attempts were run for each TSP instance of which a significant amount failed for the 6-node TSP due to runtime constraints.}
    \label{fig:powell_performance}
\end{figure*}

For the simple TSP example, the optimization process is quite successful for most optimizers, as shown in \autoref{tab:VQE_performance}. Feasibility ratios of 90\% and above are within reach, as \autoref{fig:powell_performance} illustrates while some optimizers still fail to amplify the solution state at all. Notably, the Powell, COBYLA and Nelder-Mead optimizers are stable. In high-dimensional parameter spaces such as the ones encountered in TSP, one should therefore prefer gradient-free optimizers. The Nelder-Mead method, although robust, requires a huge number of evaluations of the quantum circuit to converge and may not be suitable for practical applications. Furthermore, the number of iterations needed grows with the system size. It is worth noting that execution times on the IBM statevector simulator do not seem to depend solely on the total number of quantum circuit shots, even for the same optimizer. Scheduling might interfere here, so comparing circuit executions is a better performance metric than real time, assuming a fixed ansatz.

There is no monotone relation between the Ising energy and the application-specific performance metric $(m_{len},m_{feas})$. As seen in \autoref{fig:powell_performance} for TSP instances of size 5 and 6, there is a transition point where the feasibility increases rapidly with decreasing energy. Below this energy threshold, a high number of measured shots will be feasible. For a successful solution of the TSP, it is crucial to surpass this threshold at about 97-98\% of the optimal energy, which is far beyond QAOA's performance guarantees. The relative transition threshold depends on the penalty factor, but not the scaling. The appearance of the transition changes with the optimizer as well, see \Cref{app:cobyla} for tests with COBYLA.

Moreover, \autoref{tab:VQE_performance} includes the results of a small number of VQE runs with the best optimizers (Powell and NFT) on the 27-qubit IBM system in Ehningen. As expected, performance degrades significantly compared to the simulation, with NFT showing a better noise resilience than Powell. Only the 4-node TSP could be solved satisfactorily on real hardware.

\begin{table*}
\caption{VQE performance with various optimizers. Each row results from a set 50 individuals runs with randomly chosen initial points. The maximum number of circuit evaluations is adapted to the three hour timeout of the statevector simulator. Sets where a significant number of runs did not reach convergence within the allocated maximum number of circuit executions are marked with a star ($^*$) in the circuit executions column. Hardware runs are indicated by (H) in the optimizer column. The data is grouped by TSP size, where the analytic optimal energies for each instance are given in the intermediate rows.}
\label{tab:VQE_performance}
\begin{center}
\begin{tabular}{lcccc}
    \toprule
    Optimizer & Energy $\downarrow$ & Feas. ratio $m_{\text{feas}} \uparrow$ & TSP ratio $m_{\text{len}} \uparrow$ & Circuit executions $\downarrow$ \\
    \midrule
    \multicolumn{1}{l}{TSP size 4} & \multicolumn{2}{l}{Optimal energy $-137.72$} \\
    \midrule
    Powell & $-136.6 \pm 3.3 $ & $0.986 \pm 0.012$ & $0.95 \pm 0.12 $&$ 2000 \pm 500 $\\
    Powell (H)  & $-109 \pm 15 $& $0.37 \pm 0.34$ & $0.91 \pm 0.14$ & $900 \pm 800$ \\
    COBYLA  &$ -125 \pm16$ & $0.60 \pm 0.43$ & $0.89 \pm 0.16$ & $330 \pm 60$  \\
    Nelder-Mead & $-74.3 \pm 4.0$ & $(2.5 \pm 2.8)\cdot 10^{-3}$ & $0.85 \pm 0.13$ & $84000^*$  \\
    SLSQP &  $-13 \pm 40$ & $0.017 \pm 0.038$ & $0.91 \pm 0.12$ & $1606 \pm 25$  \\
    SPSA &  $-132.2 \pm 9.0$ & $0.84 \pm 0.37$ & $0.92\pm0.14$ & $16000^*$  \\
    NFT  & $-136.5 \pm 4.0$ & $0.9982 \pm 1.8\cdot 10^{-3}$ & $0.97 \pm 0.09$ & $100000^*$  \\
    NFT (H) & $-128.7 \pm 1.5$ & $0.947 \pm 0.032$ & $0.88 \pm 0.20$ & $2000^*$ \\
    CG  & $-73.5 \pm 2.9$ & $(2.7 \pm 3.8)\cdot 10^{-3}$ & $0.88 \pm 0.15$ & $700 \pm 100$ \\
    UMDA  & $-92 \pm 14$ & $0.08 \pm 0.27$ & $0.92 \pm 0.15$ & $70000 \pm 20000$  \\

    \midrule
    \multicolumn{1}{l}{TSP size 5} & \multicolumn{2}{l}{Optimal energy $-221.10$} \\
    \midrule
    Powell & $-217.3 \pm 2.7$ & $0.79 \pm 0.35$ & $ 0.83 \pm 0.13$ & $4000 \pm 3000$  \\
    Powell (H) & $-185.5 \pm 6.7$ & $0.014 \pm 0.034$ & $0.790 \pm 0.079$ & $500 \pm 600$ \\
    COBYLA  & $-209.7 \pm 7.1$ & $0.27 \pm 0.33$ & $0.83 \pm 0.13$ & $600 \pm 100$  \\
    Nelder-Mead & $-177.4 \pm 1.9$ & $(2.5 \pm 4.2)\cdot 10^{-4}$ & $0.838 \pm 0.094$ & $8700^*$ \\
    SLSQP &  $-3 \pm 60$ & $(0.3 \pm 1.0) \cdot 10^{-3}$ & $0.82 \pm 0.15$ & $500*$  \\
    SPSA & $-215.8 \pm 5.0$ & $0.71 \pm 0.45$ & $0.87 \pm 0.13$ & $5000^*$  \\
    NFT &  $-218.8 \pm 1.4$ & $0.9968 \pm 2.1\cdot 10^{-3}$ & $0.86 \pm 0.10$ & $29000^*$  \\
    NFT (H) & $-201.0 \pm 6.7$ & $(3.5 \pm 4.3)\cdot 10^{-3}$ & $0.928 \pm 0.046$ & $1600^*$ \\
    CG & $-176.0 \pm 2.0$ & $(2.7\pm 6.8) \cdot 10^{-4}$ & $0.82 \pm 0.13$ & $1300 \pm 300$ \\
    UMDA & $-195 \pm 23$ & $0.25 \pm 0.36$ & $0.87 \pm 0.15$ & $17000 \pm 4000$  \\
    \midrule
    \multicolumn{1}{l}{TSP size 6} & \multicolumn{2}{l}{Optimal energy $-400.47$} \\
    \midrule
    Powell  & $-390.82 \pm 3.76$ & $0.423 \pm 0.40$ & $0.80 \pm 0.12$ & $2500^*$  \\
    Powell (H) & $-349.4 \pm 6.9$ & $0.005 \pm 0.010$ & $0.75 \pm 0.13$ & $800 \pm 600$ \\
    COBYLA  & $-381 \pm 11$ & $0.185 \pm 0.27$ & $0.79 \pm 0.13$ & $924 \pm 200$  \\
    Nelder-Mead & $-350.6 \pm 2.8$ & $(1.1 \pm 2.9)\cdot  10^{-4}$ & $0.78 \pm 0.12$ & $1200^*$  \\
    SLSQP  & $12 \pm 93$ & --- & --- & $56^*$  \\
    SPSA  & $-379 \pm 12$ & $0.09 \pm 0.18$ & $0.77 \pm 0.11$ & $990^*$  \\
    NFT  & $-395.3 \pm 4.1$ & $0.94 \pm 0.24$ & $0.79 \pm 0.12$ & $3800 \pm 700$  \\
    NFT (H) & $-362.1 \pm 2.5$ & $(3.5 \pm 4.0) \cdot 10^{-4}$ & $0.734 \pm 0.070$ & $1600 \pm 100$ \\
    CG  & $-351.2 \pm 1.9$ & $(4.9 \pm 6.1)\cdot 10^{-5}$ & $0.75 \pm 0.11$ & $1300 \pm 300$  \\
    UMDA  & $-232 \pm 32$ & --- & --- & $1600 \pm 200$  \\
    \bottomrule
\end{tabular}
\end{center}
\end{table*}

\subsubsection{Penalty}
\label{sec:vqe_penalty}

The selection of suitable penalty factors is an open research question that receives constant interest. For the 4- and 5-node TSPs, we have determined that the minimum penalties $P_{min}$ required to ensure the lowest energy state is also a feasible (and therefore optimal) solution are approximately $50$ and $75$, respectively. In our simulations, we tested penalty factors ranging from approximately $P_{min}$ to $3 P_{min}$, using the best optimizers identified beforehand, NFT and Powell. The scaling factor was kept constant. The optimal energies still shift due to the constant offset between the QUBO and Ising formulations, which is dependent on the penalty.

For the 4-node TSP (data see \Cref{app:penalty_4}), the quantum state found by the NFT optimizer contains almost exclusively feasible states regardless of the penalty. The TSP length ratio seems to reach a stable plateau at about $P=70$, but is overall high. However, the optimizer uses its iteration budget (maxfev) entirely on most runs. Powell, on the other hand, converges faster, but achieves lower TSP ratios for penalties $P=50$ and $P=60$. The solution quality stabilizes again around a penalty factor of $P=70$.

The results for the 5-node TSP in \autoref{tab:VQE_penalty_5} clearly demonstrate the impact of penalties below $P_{min} \approx 75$. For $P=50$, the feasibility is very low (the uniform superposition would be at $4!/2^{16} \approx 4\cdot 10^{-4}$). The feasibility ratio at $P=70$ and $80$ rises significantly and, most interestingly, has a high variance, suggesting a feasibility phase transition when the lowest feasible state crosses the lowest infeasible one. Above this transition, NFT achieves good feasibility ratios, whereas Powell remains in a medium feasibility ratio regime with high variance between runs. Both optimizers' average TSP length ratio stays around $75-85\%$, (uniform superposition: $79\%$), however variance indicates runs with it coming close to optimality. NFT once again depletes its entire iteration budget, while Powell terminates after roughly 3700 executions of the quantum circuit (excluding shots).

\begin{table*}
\caption{Performance of VQE for various penalties for the 5-node TSP instance. $P_{min}$ is approximately 75 for this problem instance. Each row is the result of a set of 50 individual runs with randomly chosen initial points. All runs with the NFT optimizer finished due to reaching the maximum number of circuit executions (marked by $^*$).}
\label{tab:VQE_penalty_5}
\begin{center}
\begin{tabular}{cccccccc}
    \toprule
    Penalty & Opt. Energy & Optimizer  & Energy $\downarrow$ & Approx. ratio $\uparrow$ & Feas. ratio $m_{\text{feas}} \uparrow $ & TSP ratio $m_{\text{len}} \uparrow$ & Circuit executions $\downarrow$\\
    \midrule
    50 & $-143.44$ & NFT & $-142.5 \pm 1.1$ & 99.34\%& $(4 \pm 19) \cdot 10^{-3}$ & $0.03 \pm 0.15$ & $10000^*$  \\
    60 & $-157.86$ & NFT & $-156.6 \pm 1.1$ & 99.20\% & $(1.2 \pm 3.4)\cdot 10^{-3}$ & $0.29 \pm 0.40$ & $10000^*$  \\
    70 & $-172.29$ & NFT & $-171.71 \pm 0.28$ & 99.67\% & $0.36 \pm 0.48$ & $0.37 \pm 0.43$ & $10000^*$  \\
    80 & $-188.13$ & NFT & $-186.67 \pm 0.91$ & 99.22\% & $0.32 \pm 0.47$ & $0.42 \pm 0.44$ & $10000^*$  \\
    90 & $-204.61$ & NFT & $-201.8 \pm 2.2$ & 98.65\% & $0.9951 \pm 4.0\cdot 10^{-3}$ & $0.87 \pm 0.10$ & $10000^*$  \\
    100 & $-221.10$ & NFT & $-218.3 \pm 2.0$ & 98.75\%& $0.9963 \pm 2.6\cdot 10^{-3}$ & $0.85 \pm 0.10$ & $10000^*$  \\
    120 & $-254.06$ & NFT & $-251.2 \pm 2.0$ & 98.87\%& $0.9972 \pm 2.1\cdot 10^{-3}$ & $0.83 \pm 0.10$ & $10000^*$  \\
    150 & $-303.51$ & NFT & $-300.6 \pm 2.1$ & 99.03\%& $0.9967 \pm 2.2\cdot 10^{-3}$ & $0.86 \pm 0.10$ & $10000^*$  \\
    200 & $-385.92$ & NFT& $-382.1 \pm 3.2$ & 99.02\%& $0.9971 \pm 1.9\cdot 10^{-3}$ & $0.78 \pm 0.13$ & $10000^*$ \\
    \midrule
    50 & $-143.44$ & Powell & $-141.1 \pm 4.5$ & 98.35\%& $(2.3 \pm 5.7)\cdot 10^{-3}$ & $0.86 \pm 0.12$ & $2500 \pm 1400$  \\
    60 & $-157.86$ & Powell & $-153.6 \pm 7.1$ & 97.32\% & $(3.9 \pm 7.8)\cdot 10^{-3}$ & $0.88 \pm 0.12$ & $3600 \pm 1500$  \\
    70 & $-172.29$ & Powell & $-167.9 \pm 6.4$ & 97.46\% & $0.20 \pm 0.36$ & $0.74 \pm 0.11$ & $3500 \pm 1700$  \\
    80 & $-188.13$ & Powell & $-181 \pm 10$ & 96.07\% & $0.08 \pm 0.28$ & $0.81 \pm 0.12$ & $4100 \pm 1600$ \\
    90 & $-204.61$ & Powell & $-198 \pm 10$& 96.66\% & $0.54 \pm 0.48$ & $0.84 \pm 0.11$ & $4000 \pm 1700$  \\
    100 & $-221.10$ & Powell & $-212 \pm 13$ & 95.84\%& $0.58 \pm 0.48$ & $0.87 \pm 0.12$ & $3800 \pm 1300$  \\
    120 & $-254.06$ & Powell & $-240 \pm 19$ & 94.43\%& $0.69 \pm 0.43$ & $0.80 \pm 0.13$ & $3800 \pm 1700$  \\
    150 & $-303.51$ & Powell & $-286 \pm 24$ & 94.17\%& $0.64 \pm 0.47$ & $0.83 \pm 0.12$ & $3200 \pm 1200$\\
    200 & $-385.92$ & Powell & $-363 \pm 28$ & 94.05\%& $0.60 \pm 0.50$ & $0.75 \pm 0.12$ & $3700 \pm 900$ \\
    
    \bottomrule
\end{tabular}
\end{center}
\end{table*}

\section{Discussion}
\label{sec: discussion}

We have conducted an evaluation of the usefulness of current-era QC restricted to NISQ devices for the highly relevant TSP application. The TSP is a simpler variant of the CVRP, which is particularly difficult to solve using classical computing methods. To this end, we have presented a QUBO formulation of the CVRP that requires fewer qubits than previously derived formulations. Our numerical studies are based on two prominent variational algorithms for optimization problems, VQE and its special case, QAOA. Their performance was evaluated on simulators as well as in first hardware experiments. To reflect the actual solution quality of the algorithm, we have proposed a pair of performance metrics, $\left(m_{\text{len}}, m_{\text{feas}} \right)$, which measures the ratio of obtained path length to the optimal (or best known) one and the feasibility ratio, respectively. This performance  metric is necessary to account for infeasible solutions in contrast to the approximation ratio for MaxCut. It enables the user to quickly judge the quality of the algorithm from an application viewpoint and to determine whether it can yield helpful solutions to their problem. In practice, the Ising energy is not of interest to end users, but the feasibility and quality of the final path. Our experiments have shown that the Ising energy has no monotone relation with the performance metrics. Instead, at an (Ising) approximation ratio around $97-98\%$, the feasibility rises significantly and is close to zero below. This explains why QAOA failed to obtain feasible results since it stays below the threshold for all hyperparameter choices. The high approximation ratio required is partly caused by the problem formulation and might be avoided with a different encoding. In our suggested formulation, we state several constraint terms, and violating any constraint increases the energy, thus, the energy spectrum is quite wide with a large part of it only differentiating between infeasible states. Furthermore, extensions like rQAOA, WS-QAOA and QAOA with a constraint-preserving mixer are a possible path to improve the convergence process, but require more tuning and did not converge to the optimal solution with an out-of-the-box implementation either. On the other hand, hardware-efficient VQE performs well for toy instances of 4, 5 and 6 cities with very little hyperparameter tuning. In particular, Powell and NFT are powerful optimizers that can tackle the high-dimensional parameter space of 3 parameters per qubit without major problems. The parameter landscape of VQE has a simple structure, especially when compared to QAOA, cf. \autoref{fig:losslandscapes}, which enables the optimizer to find good parameters. However, the performance decreases again on real hardware which is expected due to noise and runtime constraints. With a more thorough ansatz selection and hyperparameter tuning, good results on hardware do not seem out of reach. Ans\"atze with a more favorable tuning of the parameter number are of particular interest. The ansatz used in \autoref{sec:vqe} needs three parameters per qubit, or a total of $3(n-1)^2$ parameters for a $n$-vertex TSP. 

\section{Outlook}\label{sec: outlook}

This work serves as a foundational exploration of the CVRP with variational quantum algorithms, highlighting the obstacles in obtaining useful solutions. While the study's outcomes demonstrate the difficulty of achieving high quality solutions, there are several proposals of interest for future studies, such as encoding mechanisms or algorithms. One promising approach attempts to obtain more favorable loss landscapes with alternative encodings (e.g., binary, domain-wall). Additionally, using encodings where every state corresponds to a feasible solution will increase the performance in $m_{\text{feas}}$. However, it is unlikely that algorithms based on QAOA will be able to reach the approximation ratio required for a high feasibility ratio. Still, constraint-preserving mixers might bypass this issue entirely if they can be implemented effectively on NISQ devices.
\balance
The exploration of recursive QAOA for the use case runs into the issue of imperfect correlations, even in the ideal solution state as outlined in \autoref{sec:rQAOA}. This issue goes beyond the convergence problems with QAOA in practice. It is rooted in the degeneracy of the solution state originating from the inversion symmetry of the paths (as a reminder, the cyclic permutation symmetry is eliminated in the problem formulation). There are two ways to address this problem: One could implement an additional constraint (e.g., ``city 1 is visited before city 2``) or include an intermediate processing step after each QAOA iteration, essentially mapping one of the two degenerate basis states onto the other one. The latter option requires modifying the hybrid algorithm itself whereas the first still works with the basic QAOA, but is complex to encode. To continue in this direction, constraint terms depending on the parity of a permutation may be worth exploring. Furthermore, in light of the experimental difficulties with QAOA, an analogous recursive VQE protocol could be the superior choice.
Further insights on the performance of WS-QAOA can be gained from implementations for additional problem instances of the TSP and CVRP.
In the QAOA variant with a constraint-preserving mixer, the principal tasks are to reduce the size of the mixer and examine its stability against noise. To this end, alternative encodings may be beneficial, particularly those with a higher density of feasible states in the phase space. VQE, on the other hand, has a good performance baseline even with a naive ansatz. Further efforts are necessary to estimate its scaling behavior, simplify the ansatz, promote noise resilience, and improve its convergence for larger problems. A candidate method is, e.g., filtered VQE \cite{Amaro_FVQE_2022}.

However, the roles of many parameters, such as the initial point, penalty, or scaling, remain on heuristic grounds. Therefore, additional numerical experiments and analytic efforts are necessary to expand the study to other problem instances an gain a deeper understanding of the optimization process.
\section*{Acknowledgment}
This project is supported by the Federal Ministry for Economic Affairs and Climate Action on the basis of a decision by the German Bundestag through the project \emph{Quantum-enabling Services and Tools for Industrial Applications (QuaST)}. QuaST aims to facilitate the access to quantum-based solutions for optimization problems and to bridge the gap between business and technology. Our work contributes to understanding the challenges therein for the CVRP.

\onecolumn
\clearpage
\appendix

\subsection{QAOA Performance}
\label{app:qaoa}
As described in \autoref{sec:qaoa}, QAOA generally did not achieve to produce feasible solutions. The optimal state is far from what any optimizer is able to achieve. In addition to the results shown in \autoref{tab:QAOA_performance}, different hyperparameter combinations according to \autoref{tab:hyperparams_TSP} were tested with similarly discouraging results.

\begin{table*}[h]
\caption{QAOA performance with various optimizers in simulation on the 4-city problem instance. Each row is the aggregate of 50 individual runs with a random initial point. The problem was formulated with $P=100$ resulting in an optimal energy of $-137.72$. Values marked with $^*$ indicate that a significant portion of the runs finished due to reaching the maximum allowed number of circuit executions.}
\label{tab:QAOA_performance}
\begin{center}
\begin{tabular}{lcccc}
    \toprule
    Optimizer & Energy $\downarrow$ & Feas. ratio $m_{feas}$ $\uparrow$ & TSP length ratio $m_{len}$ $\uparrow$ & Circuit executions $\downarrow$\\
    \midrule
    CG & $-11.5 \pm 6.5$ & $0.0110 \pm 6.6 \cdot 10^{-3}$ & $0.846 \pm 0.043$ & $290 \pm 50$ \\
    COBYLA & $-23 \pm 15$ & $0.014 \pm 0.011$ & $0.847 \pm 0.039$ & $97 \pm 10$  \\
    Nelder-Mead &  $-25 \pm 10$ & $0.015 \pm 0.010$ & $0.857 \pm 0.050$ & $7000^*$  \\
    NFT &  $-57 \pm 10$ & $0.021 \pm 0.017$ & $0.854 \pm 0.057$ & $19000^*$  \\
    Powell & $-48 \pm 27$ & $0.018 \pm 0.017$ & $0.868 \pm 0.061$  & $800 \pm 400$  \\
    SPSA & $-45.9 \pm 9.5$ & $0.018 \pm 0.018$ & $0.857 \pm 0.050$ & $10000^*$ \\
    SLSQP &  $-2.0 \pm 6.7$ & $0.0112 \pm 8.5\cdot 10^{-3}$ & $0.849 \pm 0.049$ & $12 \pm 4$  \\
    UMDA & $-57.3 \pm 7.7$ & $0.023 \pm 0.022$ & $0.850 \pm 0.049$ & $5000 \pm 1000$\\
    \bottomrule
\end{tabular}
\end{center}
\end{table*}

\subsection{VQE with the COBYLA optimizer}
\label{app:cobyla}
\begin{figure*}[h]
    \centering
    \includegraphics[width=\linewidth]{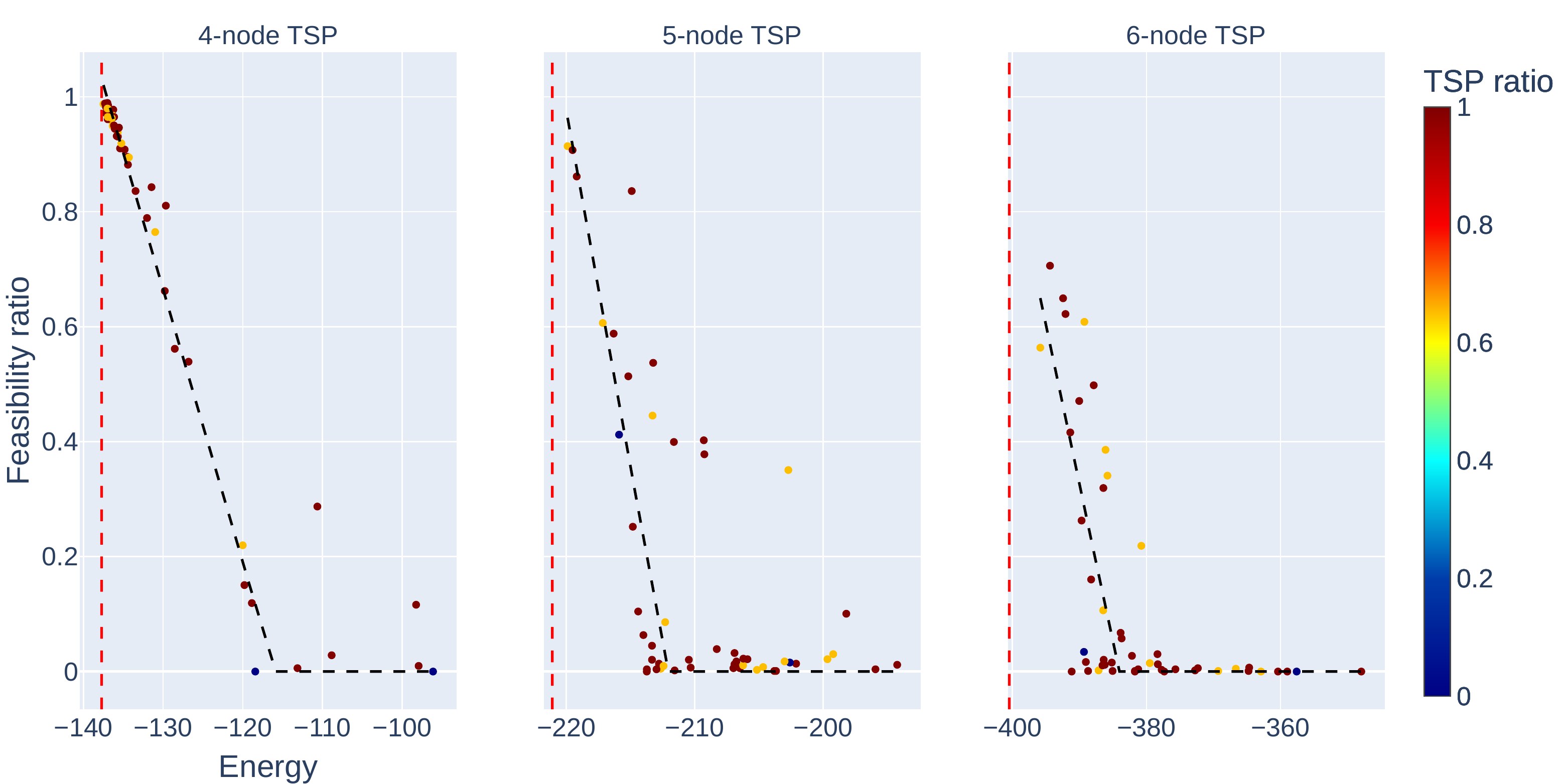}
    \caption{The relation of the feasibility ratio to the Ising energy of VQE with the COBYLA optimizer on a TSP with 4 to 6 nodes. In contrast to Powell, COBYLA seems to show a linear increase in feasibility below a threshold energy. 50 attempts were run for each TSP instance.}
    \label{fig:cobyla_performance}
\end{figure*}

In the same manner as \autoref{fig:powell_performance}, optimizing the VQE ansatz with COBYLA as shown in \autoref{fig:cobyla_performance} shows an energy threshold below which feasible states start to appear. However, the relation between $m_{feas}$ and the Ising energy at the transition point seems different. Powell exhibits an approximate step-like behaviour, modeled in \autoref{fig:powell_performance} through an arcus tangens. For COBYLA, it looks like the feasibility ratio increases linearly below a threshold energy. Further research is needed in this direction, but this phenomenon hints towards the best optimizer choice depending on the specific application as well as the formulation and encoding of the problem. Different optimizers may take different paths towards the optimum, resulting in approximate results of different quality. Once again, this is why one should consider application-specific performance metrics and cannot rely purely on energy approximation ratios.

\subsection{Influence of the Penalty Factor for the 4-node TSP instance}
\label{app:penalty_4}
Results with varying penalties for the 4-node TSP instances are shown in \autoref{tab:VQE_penalty_4}. The NFT and Powell optimizers reach approximation ratios around $98-99\%$ for all penalty factors. Likewise, the feasibility and TSP length ratios obtained are similarly well, with the exception that Powell starts to reach high feasibility only starting at $P=70$. Furthermore, with penalties only slightly higher than $P_{min}$, the performance metrics fluctuate more and stabilizes around $P=70 \approx 1.4P_{min}$ for both optimizers. Within the tested values from $P_{min}=50$ to $4P_{min}=200$, no significant performance impact can be observed. . NFT still uses its budget of quantum circuit evaluations completely. A modification of the convergence criterion might be useful for this optimizer. Powell requires significantly fewer executions for a feasibility ratio that is only marginally lower.

\begin{table*}
\caption{Performance of VQE for various penalties for the 4-node TSP instance. $P_{min}=50$ is the approximate minimum penalty for the 4-node TSP. All simulations were given a budget of 10000 quantum calls. The overall scaling was held constant, but different penalty factors still result in varying optimal energies}
\label{tab:VQE_penalty_4}
\begin{center}
\begin{tabular}{lccccccc}
    \toprule
    Penalty & Opt. Energy & Optimizer  & Energy $\downarrow$ & Approx. ratio $\uparrow$ & Feas. ratio $m_{feas}$ $\uparrow$ & TSP length ratio $m_{len}$ $\uparrow$  & Circuit evaluations $\downarrow$\\
    \midrule
    $50$ & $-80.90$ & NFT & $-80.69 \pm  0.93$ & $99.74\%$& $0.98 \pm 0.14$ & $0.91 \pm 0.15$ & $10000^*$  \\
    $60$ & $-92.27$ & NFT & $-91.9 \pm 2.0$ & $99.60\%$ & $0.98 \pm 0.14$ & $0.969 \pm 0.095$ & $10000^*$  \\
    $70$ & $-103.62$ & NFT & $-103.2 \pm 2.4$ & $99.59\%$ & $0.99836 \pm 6.4\cdot 10^{-4}$ & $0.976 \pm 0.083$ & $10000^*$  \\
    $80$ & $-114.99$ & NFT & $-113.6 \pm 4.6$ & $98.79\%$ & $0.99834 \pm 6.6\cdot 10^{-4}$ & $0.96 \pm 0.10$ & $10000^*$ \\
    $90$ & $-126.36$ & NFT & $-125.9 \pm 2.4$ & $99.64\%$ & $0.99832 \pm 6.4\cdot 10^{-4}$ & $0.976 \pm 0.083$ & $10000^*$ \\
    $100$ & $-137.72$ & NFT & $-135.9 \pm 5.1$ & $98.68\%$& $0.98 \pm 0.14$ & $0.96 \pm 0.11$ & $10000^*$  \\
    $120$ & $-160.45$ & NFT & $-158.6 \pm 5.1$ & $98.85\%$& $0.98 \pm 0.14$ & $0.981 \pm 0.070$ & $10000^*$ \\
    $150$ & $-194.54$ & NFT & $-194.1 \pm  2.4$ & $99.77\%$& $0.99849 \pm 6.1\cdot 10^{-4}$ & $0.983 \pm 0.069$ & $10000^*$  \\
    $200$ & $-251.35$ & NFT& $-249.9 \pm 4.6$ & $99.42\%$& $0.99846 \pm 6.4\cdot 10^{-4}$ & $0.99713 \pm 9.7 \cdot 10^{-4}$ & $10000^*$ \\
    \midrule
    $50$ & $-80.90$ & Powell & $-79.9 \pm 2.1$ & $98.76\%$& $0.65 \pm 0.46$ & $0.88 \pm 0.16$ & $2000 \pm 600$  \\
    $60$ & $-92.27$ & Powell & $-90.4 \pm 4.7$ & $98.00\%$ & $0.87 \pm 0.33$ & $0.89 \pm 0.16$ & $2000 \pm 600$ \\
    $70$ & $-103.62$ & Powell & $-102.0 \pm 4.7$ & $98.51\%$ & $0.93 \pm 0.24$ & $0.95 \pm 0.12$ & $1900 \pm 500$  \\
    $80$ & $-114.99$ & Powell & $-113.8 \pm 4.1$ & $98.96\%$ & $0.9923 \pm 6.5\cdot 10^{-3}$ & $0.983 \pm 0.070$ & $2000 \pm 600$ \\
    $90$ & $-126.36$ & Powell & $-125.8 \pm 2.4$ & $99.59\%$ & $0.9907 \pm 9.2\cdot 10^{-3}$ & $0.99 \pm 0.13$ & $2000 \pm 600$  \\
    $100$ & $-137.72$ & Powell & $-137.2 \pm 2.5$ & $99.63\%$& $0.9920 \pm 5.2\cdot 10^{-3}$ & $0.99 \pm 0.11$ & $2100 \pm 600$  \\
    $120$ & $-160.45$ & Powell & $-158.5 \pm 7.4$ & $98.77\%$& $0.95 \pm 0.20$ & $0.95 \pm 0.12$ & $2000 \pm 600$  \\
    $150$ & $-194.54$ & Powell & $-193.6 \pm 3.3$ & $99.53\%$& $0.9919 \pm 6.1\cdot 10^{-3}$ & $0.990 \pm 0.049$ & $2000 \pm 600$  \\
    $200$ & $-251.35$ & Powell & $-250.3 \pm 3.4$ & $99.58\%$& $0.991 \pm 0.011$ & $0.990 \pm 0.051$ & $2100 \pm 600$  \\
    
    \bottomrule
\end{tabular}
\end{center}
\end{table*}

\clearpage
\twocolumn
\printbibliography

\end{document}